\documentclass{article}
\usepackage{arxiv}
\usepackage[utf8]{inputenc} 
\usepackage[T1]{fontenc}    
\usepackage{hyperref}       
\usepackage{url}            
\usepackage{booktabs}       
\usepackage{amsfonts}       
\usepackage{nicefrac}       
\usepackage{microtype}      
\usepackage{lipsum}
\usepackage{graphicx}
\usepackage{multirow}
\usepackage{float}
\usepackage{rotating}
\usepackage{subcaption}
\usepackage{xcolor}
\usepackage{amsmath} 
\usepackage{setspace}
\usepackage[numbers]{natbib}
\usepackage{algorithmic}
\usepackage{amsmath}
\usepackage{amsfonts}

\usepackage{threeparttable}
\usepackage{booktabs}
\usepackage{multirow}
\usepackage{xcolor}
\usepackage{colortbl}

\definecolor{salmon}{RGB}{247, 173, 150}
\newcommand{\percentbar}[1]{%
  \begin{minipage}{2cm}
    \colorbox{salmon}{\makebox[\fpeval{#1/100}\linewidth][l]{\phantom{text}}} \hspace{0mm} #1\%
  \end{minipage}
}
\title{Characterizing AI Manipulation Risks \\
in Brazilian YouTube Climate Discourse}

\author{
 Wenchao Dong \\
  MPI-SP \\
   \And
   Marcelo Sartori Locatelli \\
  MPI-SP, UFMG \\ \\
  \And
  Virgilio Almeida \\
  UFMG \\
  \And
 Meeyoung Cha \\
 MPI-SP, KAIST \\
}

\begin{document}
\maketitle

\begin{abstract}
Climate change poses a global threat to public health, food security, and economic stability.
Addressing it requires evidence-based policies and a nuanced understanding of how the threat is perceived by the public, particularly within visual social media, where narratives quickly evolve through voices of individuals, politicians, NGOs, and institutions. 
This study investigates climate-related discourse on YouTube within the Brazilian context, a geopolitically significant nation in global environmental negotiations. Through three case studies, we examine (1) which psychological content traits most effectively drive audience engagement, (2) the extent to which these traits influence content popularity, and (3) whether such insights can inform the design of \textit{persuasive} synthetic campaigns—such as climate denialism—using recent generative language models. 
Another contribution of this work is the release of a large publicly available dataset of 226K Brazilian YouTube videos and 2.7M user comments on climate change.\footnote{Zenodo: \url{https://doi.org/10.5281/zenodo.17551955}} The dataset includes fine-grained annotations of persuasive strategies, theory-of-mind categorizations in user responses, and typologies of content creators. This resource can help support future research on digital climate communication and the ethical risk of algorithmically amplified narratives and generative media.

\end{abstract}

\section{Introduction}

\begin{figure*}[t]
    \centering
    \includegraphics[width=\linewidth]{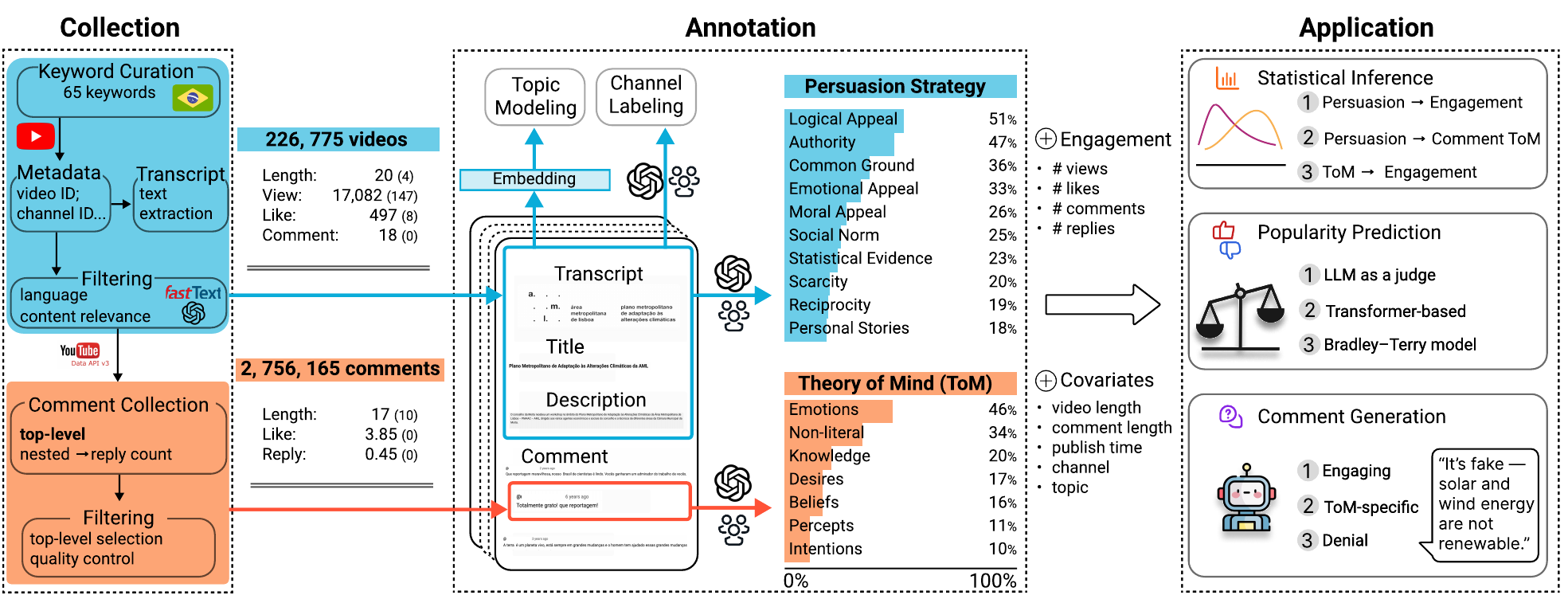}
    \caption{Proposed analytical framework includes data collection and preprocessing, annotations, and three case studies to understand the user engagement. Videos and comments' mean values, with median values in parentheses, are included. Ten persuasion strategies and seven theory of mind categories are ordered based on the distributions over the entire dataset.}
    \label{fig:pipeline}
\end{figure*}

The pursuit of collective peace and prosperity among nations has long been a key objective on a global scale. One embodiment of such ideals is the United Nations' Sustainable Development Goals (SDGs), which call on countries to tackle pressing global issues through a shared blueprint. Among them is the SDG~\#13, or ``climate action,'' which has received considerable attention, particularly as extreme weather events become more frequent~\cite{national2016attribution}, posing a threat to public health, food security, and economic stability~\cite{solomon2019climate}. In this context, understanding climate change discourse is increasingly important, as it can support efforts to increase public awareness on policies that address societal needs.

Understanding public discourse around climate change remains critical, particularly as social media platforms increasingly shape how climate narratives are communicated and received—subtly influencing public sentiment and policy conversations.
Brazil plays a prominent role in these discussions, not only as a leading nation in the Global South, but also as a host of major climate summits such as the 30th United Nations Framework Convention on Climate Change (UNFCCC), or COP30. Its influence is further amplified by being home to a substantial portion of the Amazon rainforest and more than 10\% of the world’s freshwater resources. Together, these factors position Brazil as a key focal point for climate change-related research and policy development.

To characterize climate discourse in Brazil, we collected a large-scale dataset of Brazilian Portuguese YouTube content spanning 2019 to 2025, with a total of 226,775 videos and 2,756,165 user comments. As of 2025, YouTube reached approximately 68\% of the Brazilian population, representing a significantly high penetration rate, and 86\% of the country's 200 million residents are connected to the Internet~\cite{DataReportal2025Brazil}.
This high level of digital engagement offers a valuable lens into how climate information is communicated and received by the public.

Building on recent advances in psychology-informed computational approaches to human decision-making~\cite{binz2025foundation}, we apply psycholinguistic methods to examine how climate information is conveyed and received on YouTube.
To characterize creator messaging, we annotate video content using 10 persuasion strategies, including emotional appeals that evoke fear or empathy, and logical arguments emphasizing cause-and-effect relationships.
To infer potential mental states underlying viewer responses, we classify user comments according to seven Theory of Mind (ToM) categories. These include, for example, expressions of intentionality toward climate-related actions (\textit{intention}) and beliefs about the legitimacy of climate change (\textit{belief}). 
Drawing on these psychological traits, we present three case studies that (1) identify key influences shaping the Brazilian climate discourse, (2) assess the predictability of content popularity, and (3) examine how these insights can inform the automatic generation of persuasive messages. The full data analysis pipeline is illustrated in Figure~\ref{fig:pipeline}.

Our study reveals that psychological features of climate discourse strongly influence audience engagement (i.e., views, comments) in Portuguese-language climate videos from Brazil. Emotionally framed messages consistently generated higher levels of user interaction, while logically or statistically driven content attracted comparatively less interaction.
These dynamics raise concerns about the potential misuse of generative technologies such as large language models (LLMs) to produce \textit{persuasive} synthetic comments. This generative capabilities can facilitate effortless reproduction of misleading narratives, including content promoting climate denialism.\footnote{Disclaimer:~All examples are illustrative and our research framework is not intended to promote or amplify misinformation.}
Given the rising concerns over synthetic consensus in generative media~\cite{schroeder2025malicious}, our study offers a psycholinguistic framework and annotated dataset to assess climate opinion manipulation in Brazil.

\section{Related Work}

\subsection{Discourse Analysis}

Social media plays a role in raising public awareness of climate issues~\cite{mavrodieva2019role}, but it can also be used to spread misinformation that fuels climate skepticism, denial, and contrarianism~\cite{treen2020online}.
One study conducted a thematic analysis of TikTok videos and found that content creators tend to mention the topic through novel lenses, such as environment-conscious lifestyle~\cite{galdeman2025mapping}.
\citet{chen2023climate} compared Twitter and news media to capture their different narrative styles.
Moreover, multiple studies have explored how climate change features in political debates, suggesting emerging patterns of polarization~\cite{shapiro2015more, jang2015polarized, falkenberg2022growing}.
As for online information validity, research has documented that discussions are often dominated by a small group of highly active users~\cite{shapiro2018climate}, which may further contribute to the spread of conspiracy theories~\cite{allgaier2019science}.

\subsection{Persuasion and Belief Adoption}

Persuasion has long been a focus of social psychology and explains how decision-making can be shaped by external influences~\cite{crano2006attitudes}.
\citet{cialdini2001science} identified key factors, such as authority and social norms, that significantly affect individual choices.
These mechanisms have proven effective in domains like marketing~\cite{kumar2023persuasion} and charity~\cite{wang2019persuasion}.
More recently, studies have highlighted the persuasive potential of AI-generated text~\cite{breum2024persuasive, salvi2025conversational}.
Experiments show that LLM-written information can influence belief formation related to conspiratorial narratives~\cite{costello2024durably, hackenburg2025levers}.
This persuasive rhetoric raised further concerns about the tailored messages that are referred to as microtargeting~\cite{tappin2023quantifying, hackenburg2024evaluating}.
Conversely, automated persuasion has also been considered a tool to reduce skepticism about change~\cite{czarnek2025addressing}.

\subsection{Theory of Mind (ToM)}

This concept was first introduced by~\citet{premack1978does} as the ability to impute mental states to themselves and others and was later identified as a core component of human cognition~\cite{leslie2004core}.
Different categories of ToM are used to assess children's cognitive abilities~\cite{lane2010theory, beaudoin2020systematic}.
Recognizing others' mental states enables the understanding of abstract concepts like beliefs in others and hence helps with decision-making~\cite{frith2008role}. 
Recent work has extended ToM to language models~\cite{ma2023towards, shapira2024clever} and demonstrated improvements in planning and reasoning performance~\cite{jung2024perceptions, cross2025hypothetical, kim2025hypothesis}.
In particular, LLMs have shown the ability to track others' mental states on par with humans~\cite{kosinski2024evaluating, strachan2024testing}.

\section{Data Methodology}

\subsection{Data Collection}

Before collecting climate-related data, we reviewed existing datasets in other languages to identify common keywords and methodologies. Most were English-language collections, primarily used in studies on climate misinformation and stance detection. A summary of relevant datasets is provided in Appendix C. Building on this foundation, we assembled a large dataset incorporating psychological attributes from \emph{persuasion} and \emph{Theory of Mind (ToM)} research, aiming to capture a broader range of mental states~\cite{premack1978does}.

Based on this literature survey, we selected 65 keywords that are commonly used as search queries to gather climate-related content~\cite{salmi2022mudanccas,baltasar2022analysis} and used them for YouTube Data API v3.
We set the preferred language and location to `Portuguese' and `Brazil' for data collection, filtering out non-Portuguese videos.
To remove irrelevant content that arises from keyword ambiguity, we used GPT-4.1-mini with a temperature of 0 to exclude videos with low climate relevance.
We then collected all available transcripts and comments from the filtered videos.
For live-streamed videos with comments that were published before the video itself, we adjusted the comment timestamps to match the video's publish time to better align engagement within the streaming period.
Since nested comments often diverge from the video content~\cite{cakmak2024bias}, we included only top-level comments that are direct replies to video in our analysis. 
Further data descriptions are in~\cite{dong25goodit}.
Since YouTube engagement can vary with video length, we adopt a similar approach to~\cite{violot2024shorts} and define short videos as those under 3 minutes, based on the platform's official length threshold~\cite{youtube2024shorts}.

\begin{figure}[t]
    \centering
    \includegraphics[width=.6\linewidth]{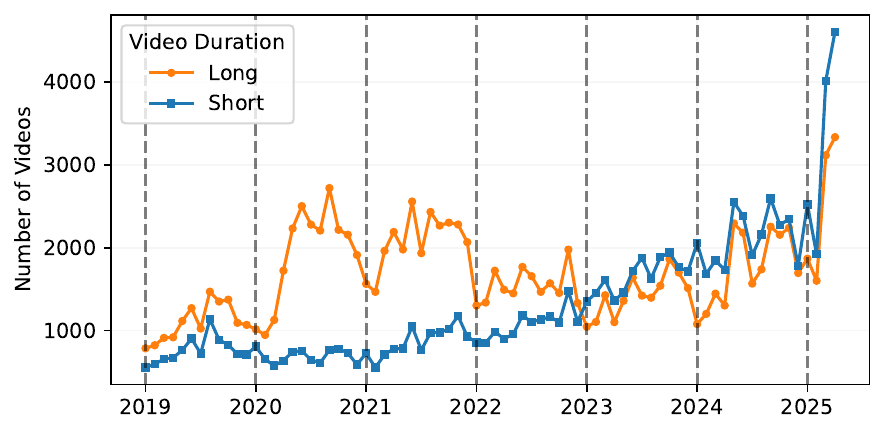}
    \caption{Monthly counts of  climate-related Brazilian videos during 2019-2025. Vertical lines mark the start of each year.}
    \label{fig:short_long_distribution}
\end{figure}

Our dataset covers a seven-year period (2019–2025) and comprises 226,775 YouTube video metadata and 2,756,165 user comments.
Figure~\ref{fig:short_long_distribution} shows the number of new video posts over time. The number of long videos (i.e., those over 3 minutes) increased during the COVID-19 pandemic (between 2020 and 2022), but was soon overtaken by the rise of short-form videos, which have steadily gained traction among creators. Since 2023, the majority of videos on the climate change topic have been shared in short-form format.

\subsection{Data Annotation}

We use the following notation to describe our dataset. Let $\mathcal{D}$ denote the complete dataset, consisting of 
$V = \{v_1, v_2, \ldots, v_n\}$, the set of $n$ videos, and $C = \{c_1, c_2, \ldots, c_m\}$, the set of $m$ comments.
We define a video-comment mapping  $\phi: C \rightarrow V$, where $\phi(c_k) = v_i$ indicates that comment $c_k$ belongs to video $v_i$.
For each video $v_i$, let $C_i = \{c_k \in C : \phi(c_k) = v_i\}$ be the set of all comments linked to video $v_i$.

\subsubsection{Persuasion Traits}

To study how creators communicate the topic to their audience, we analyze psychological linguistic patterns, focusing on 10 persuasive strategies drawn from the literature~\cite{cialdini2001science,kumar2023persuasion, costello2024durably}:
\begin{itemize}
\item \emph{Logical Appeal}: appealing with reasons and evidences
\item \emph{Emotional Appeal}: eliciting emotional feelings
\item \emph{Statistical Evidence}: providing concrete data, statistics
\item \emph{Social Norm}: creating pressure through social acceptance
\item \emph{Authority}: citing experts, institutions, and official reports
\item \emph{Personal Stories}: explaining individual experiences
\item \emph{Moral Appeal}: appealing with ethical responsibility 
\item \emph{Reciprocity}: emphasizing mutual benefits of giving back
\item \emph{Scarcity}: presenting limited time and irreversible impacts
\item \emph{Common Ground}: building shared identity and values
\end{itemize}

Language persuasiveness has received increasing attention since the emergence of LLMs~\cite{breum2024persuasive, salvi2025conversational, bai2025llm}. These models can be instructed to employ specific persuasion techniques during text generation~\cite{shi2021refineimitate,costello2024durably}. 
Given that LLMs can readily identify such persuasion strategies in text~\cite{josegreenstadt2025llms}, we used GPT-4.1 to annotate the presence of persuasion strategies in video content. All inferences were performed using five-shot prompting with a temperature setting of 0. 
To validate quality, we conducted a human evaluation and comparison, yielding an average F1 score of 0.93 and average accuracy of 0.98.
For each video, we define $\mathbf{p}_i = (p_{i,1}, p_{i,2}, \ldots, p_{i,P}) \in \mathbb{R}^{P}$ as a vector representing the presence of persuasion strategies $P$, where $p_{i,j}$ represents the presence of the persuasion strategy $j$ in the video $v_i$. 

\subsubsection{Theory of Mind (ToM) Traits}

Another key psychological dimension we examine is rooted in ToM, the capacity to understand and interpret the mental states of others. Based on literature~\cite{haznitrama2025methodologies}, we consider a wide range of ToM categories, which  serve as analytical anchors to infer potential audience reactions to climate-related videos from a third-person perspective. In this study, we evaluate 7 distinct types of ToM, based on a widely recognized taxonomy~\cite{beaudoin2020systematic, ma2023towards}: 
\begin{itemize}
    \item \emph{Belief}: information states that people hold to be true
    \item \emph{Intention}: committed choices with planned actions
    \item \emph{Desire}: motivational states representing preferences
    \item \emph{Emotion}: affective states emotional responses
    \item \emph{Knowledge}: organized representations of information
    \item \emph{Percept}: sensory or socially shared perceptions 
    \item \emph{Non-literal}: using figurative or indirect language
\end{itemize}

We used GPT-4.1-mini to annotate ToM types via a five-shot prompting approach with a temperature of 0. To validate the annotations, we applied two methods: (1) randomly sampling comments and re-annotating them with a larger language model (e.g., GPT-4.1) to assess consistency and quality; and (2) manual annotation and comparison against GPT-4.1-mini results, which yielded an average F1 score of 0.66 and average accuracy of 0.83.
For each comment $c_k$, we define $\mathbf{t}_k = (t_{k,1}, t_{k,2}, \ldots, t_{k,T}) \in \mathbb{R}^{T}$ as a vector of $T$ ToM labels, where $t_{k,j}$ denotes the presence of the ToM aspect $j$ in comment $c_k$.

\subsubsection{Topic and Channel Modeling}

To better control for factors influencing video engagement~\cite{dong25goodit}, we apply topic clustering and channel annotation. 
We cluster video content $V$ into thematic groups using the unsupervised BERTopic algorithm~\cite{grootendorst2022bertopic}.
Channel labels were generated using GPT-4.1 through iterative prompt refinement and subsequently validated against human annotations (see Appendix F).

\begin{figure*}[t]
    \centering
    \hspace*{-6mm}
    \includegraphics[width=1.05\linewidth]{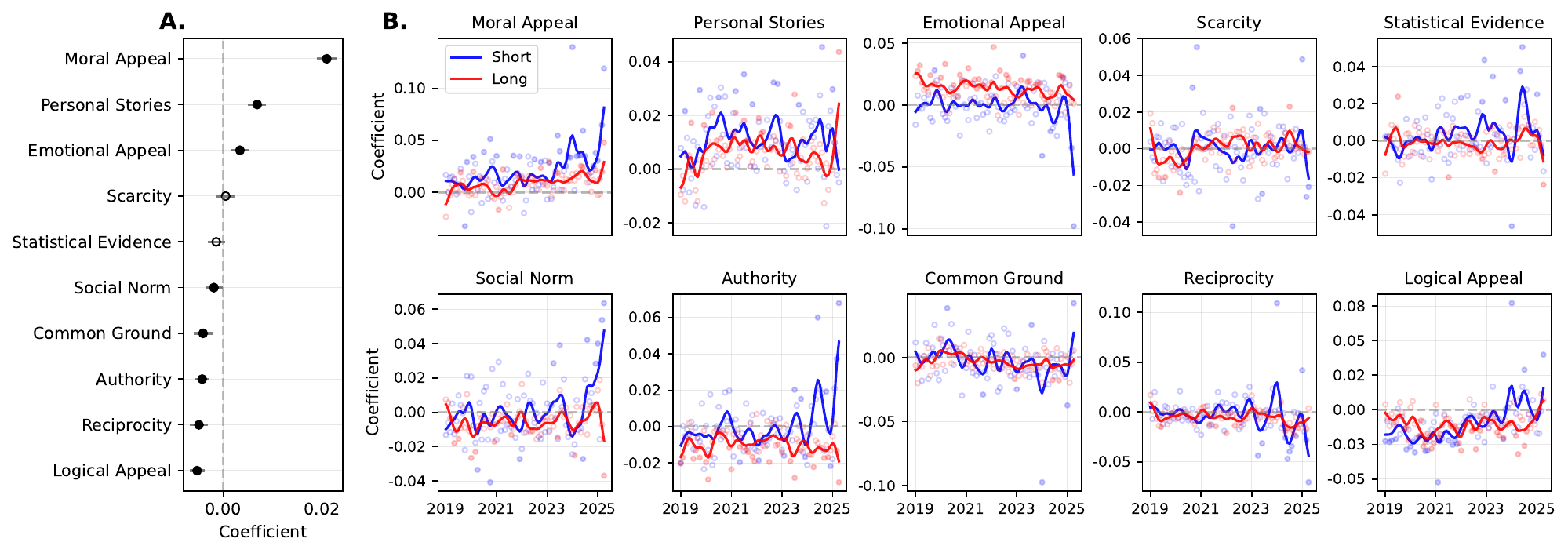}
    \vspace*{-3mm} 
    \caption{Effect of different persuasion strategies on the like ratio for (A) all videos and (B) monthly trends by video length. Solid points indicate statistically significant regression coefficients at the 0.05 level.}
    \label{fig:persuasion_regression}
\end{figure*}

\begin{figure}[t]
    \centering
    \includegraphics[width=0.8\linewidth]{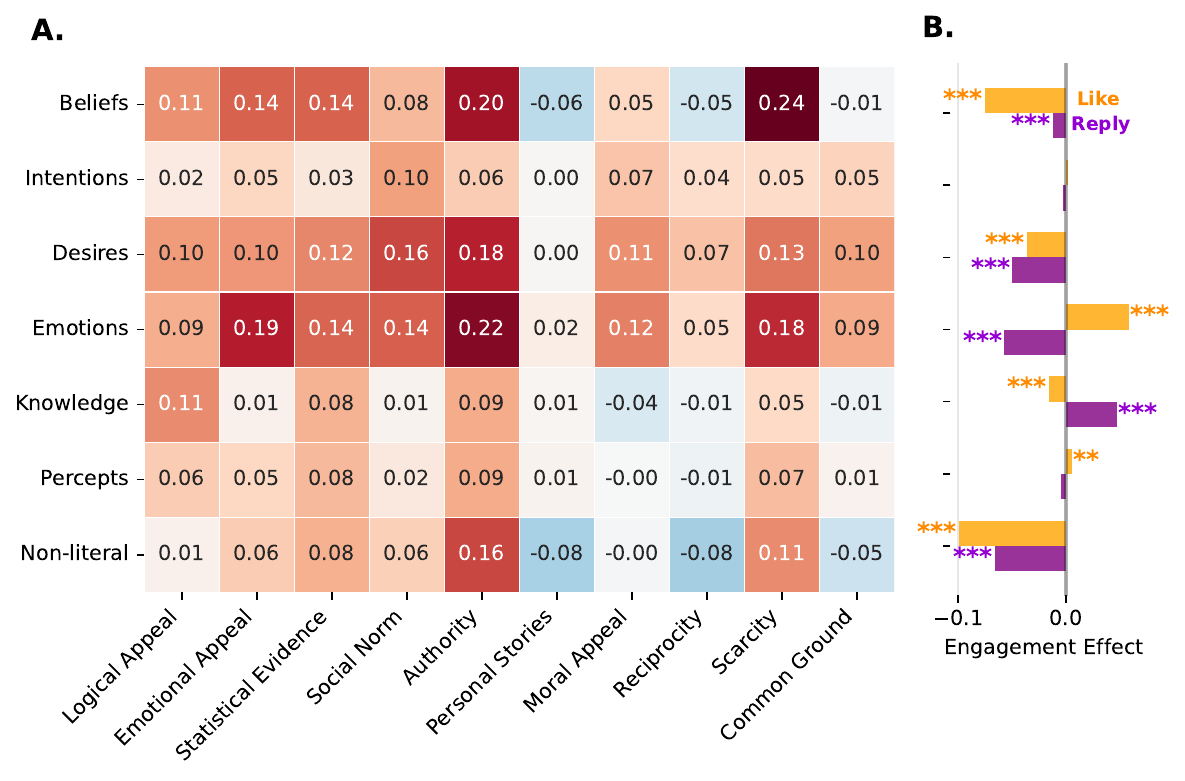}
    \caption{(A) Pearson correlations at the video level between 10 persuasion strategies and 7 ToM categories. (B) Effects of ToM mental states on audience engagement of likes and replies.  \textit{Note.} $^{*}p < .05$; $^{**}p < .01$; $^{***}p < .001$}
    \label{fig:theory_of_mind}
\end{figure}

\section{Discourse Manipulation}

Building on the methodology above, we present three case studies that collectively suggest the potential risk of discourse manipulation using generative AI.

\subsection{Case Study 1:  Engagement Modeling}

The first case study examines the relationship between psychological traits and audience engagement at two levels. The first is \emph{video-level engagement}, defined by the number of likes and comments per video. The second is \emph{comment-level engagement}, defined by the number of likes and replies per individual comment.
To account for differences in view counts, we normalize the number of likes and comments by the number of views for each video $v_i$ and analyze the \textit{like ratio} $L_i$ and \textit{comment ratio} $R_i$ instead. For each comment $c_k$, we denote the number of likes and replies as $\ell_k$ and $r_k$.

\subsubsection{Psycholinguistic Traits}

We assess how psycholinguistic narratives in climate discourse influence user engagement through a three-stage evaluation process.
First, we quantify the influence of persuasion strategies $\mathbf{p}_i$ on video-level metrics $L_i$ and $R_i$ using linear regression models. These models control for confounding factors such as video length and the publishing channel. 
Next, we further validate the effects of persuasion across the temporal span of our dataset and examine whether the conclusions hold across different video lengths. As a robustness check, we replicate these analyses within each discussion topic and channel.

Second, since each video contains multiple comments, we aggregate Theory of Mind (ToM) categories across all relevant comments and align the resulting summary with the video-level persuasion score.
For each video $v_i$, we compute the aggregated ToM vector as $\bar{\mathbf{t}}_i = (1/ |C_i|) \sum_{c_k \in C_i} \mathbf{t}_k$, where $|C_i|$ is the number of comments associated with $v_i$.
We then compute the correlations between $\mathbf{p}_i$ and $\bar{\mathbf{t}}_i$, controlling for video length and channel subscriber count.

Third, user engagement at the comment level is modeled using regression, with the number of likes ($\ell_k$) and replies ($r_k$) for each comment as the dependent variable, and the ToM score ($\mathbf{t}_k$) as the independent variable. Additional control variables include comment length, publishing channel, and the temporal gap between video publication and comment posting time.

\subsection{Case Study 2: Popularity Prediction}

The next case study evaluates how these traits influence video popularity, a key factor in assessing the feasibility of automated information manipulation~\cite{wu2018beyond}.
In the context of YouTube video comments, for any pair of comments $(c_i, c_j)$, we define the relative likes preference $y_{ij}^{(\ell)}$ as a binary indicator of whether comment $c_i$ received more likes than comment $c_j$: $y_{ij}^{(\ell)} = \mathbb{I}[\ell_i > \ell_j]$, where $\mathbb{I}[\cdot]$ denotes the indicator function.
The overall accuracy is computed as the average prediction correctness for each pair $A_{ij} = \mathbb{I}[y_{ij}^{(\ell)} = \hat{y}_{ij}^{(\ell)}]$, where $\hat{y}_{ij}^{(\ell)}$ is a model's prediction.

We evaluate the extent to which popularity is influenced by ToM.
For each trait $j$, we partition the comment pairs based on whether the more popular comment $c_i$ exhibits the trait ($\mathcal{P}_j^{+}$) or not ($\mathcal{P}_j^{-}$), given that ToM traits tend to be more prevalent among popular comments.
We then compute the difference in prediction accuracy between the two sets: 
\begin{equation}
(1/|\mathcal{P}_j^{+}|) \sum_{(c_i, c_j) \in \mathcal{P}_j^{+}} A_{ij} 
- (1/|\mathcal{P}_j^{-}|) \sum_{(c_i, c_j) \in \mathcal{P}_j^{-}} A_{ij}
\noindent
\end{equation}
We lastly design multiple experimental conditions to constrain data sampling under varying levels of contextual information.

\subsubsection{Computational Models}

We use three methods to predict the relative popularity of video comments.
First, we adopt LLM-as-a-judge framework~\cite{zheng2023judging}, using five models: OpenAI GPT-4.1 and o4-mini, Microsoft Phi 4; Meta Llama-3.1-8B and Llama-4-Maverick.
Second, we fine-tune encoder-only pretrained language models: the Brazilian Portuguese BERTimbau~\cite{souza2020bertimbau} and multilingual DeBERTa V3~\cite{he2021debertav3}.
Third, we implement the Bradley-Terry model~\cite{bradley1952rank, ye2025generative}.
In this method, we compute for each comment $c_k$ an embedding representation $\mathbf{e}_k = \text{Embed}(c_k) \in \mathbb{R}^d$ and train a linear classifier to estimate the relative popularity score. The probability that comment $c_i$ is more popular than comment $c_j$ is defined as follows, with a threshold of 0.5 used to binarize the probabilities for consistent comparisons:
%
\begin{equation}
    \frac{\exp(\beta_{i,\ell})}{\exp(\beta_{i,\ell}) + \exp(\beta_{j,\ell})}, \quad \text{where} \quad \beta_{i,\ell} = \text{Linear} (\mathbf{e}_i)
\end{equation}

\subsection{Case Study 3: Comment Generation}

The final case study demonstrates how persuasive climate-related comments can be synthetically generated by leveraging the patterns identified earlier as a hypothetical risk scenario. We follow the experimental setup in~\citet{ye2025measuring} to fine-tune Llama-3-8B~\cite{grattafiori2024llama} for comment generation.
To estimate the quality of a generated comment $c_{\text{gen}}$, we retrieve $K$ most semantically similar comments from $C_{\text{ref}}$ using cosine similarity.
%
\begin{equation}
\begin{aligned}
    \hat{\ell}_{\text{gen} | K}  &= \frac{1}{K} \sum_{c' \in \mathcal{N}_K(c_{\text{gen}})} \ell_{c'}, \quad \text{where} \\
    \mathcal{N}_K(c_{\text{gen}}) &= \arg\max_{S \subseteq C_{\text{ref}}, |S|=K} \sum_{c' \in S} \cos(\mathbf{e}_{c_{\text{gen}}}, \mathbf{e}_{c'})
\end{aligned}
\end{equation}

\subsubsection{Targeted Profiles}

We construct fine-tuning datasets in three scenarios and, for each, train two model variants: a \textit{likable model} trained on the top 10\% of most liked comments, and a \textit{baseline model} trained on randomly selected comments.
The full set of comments under each scenario after excluding training samples is used as a reference set for similarity evaluations.

First, we sample videos categorized by distinct persuasion strategies to control for video-level effects while maintaining content diversity.
Second, we sample comments that match a target ToM profile $\mathbf{t}_{\text{target}} \in \mathbb{R}^T$ to generate comments that reflect specific mental states.
Third, we further subcategorize the ``Belief'' ToM category into distinct stances, including climate change belief and climate change denial.
To simulate extreme scenarios, we filter denial-related comments using targeted keywords and fine-tune an \textit{extreme denial model} capable of generating content aligned with strong climate skepticism.

\begin{table}[!h]
\small
\centering

\resizebox{\textwidth}{!}{%
\begin{tabular}{lllllllll} \toprule
                            & \textbf{GPT-4.1} & \textbf{o4-mini} & \textbf{LLaMA-3.1} & \textbf{LLaMA-4} & \textbf{Phi-4}          & \textbf{BERTimbau}      & \textbf{DeBerta V3}     & \textbf{Bradley--Terry }\\ \midrule
\multicolumn{5}{l}{\textbf{Rand}}                                                                         &                &                &                &                \\
\quad Base   & 0.75$_{.040}$    & 0.69$_{.085}$    & 0.57$_{.037}$      & 0.68$_{.072}$    & 0.72$_{.054}$  & 0.88$_{.001}$  & 0.81$_{.013}$  & 0.78$_{.002}$  \\
\quad +CO   & 0.76$_{-.010}$   & 0.73$_{.115}$    & 0.54$_{-.012}$     & 0.70$_{.112}$    & 0.68$_{.016}$  & 0.86$_{.015}$  & 0.84$_{.031}$  & 0.73$_{.020}$  \\
\quad +FS    & 0.77$_{.035}$    & 0.72$_{.091}$    & 0.63$_{.034}$      & 0.71$_{.039}$    & 0.73$_{.063}$  &         -       &   -             &   -             \\
\quad +CO+FS & 0.80$_{.031}$    & 0.72$_{.130}$    & 0.57$_{.048}$      & 0.70$_{.080}$    & 0.70$_{.086}$  &    -            &      -          &   -             \\\midrule
\multicolumn{5}{l}{\textbf{Vid.}}                                                                          &                &                &                &                \\
\quad Base   & 0.77$_{.005}$    & 0.73$_{.024}$    & 0.59$_{-.013}$     & 0.69$_{.045}$    & 0.74$_{.022}$  & 0.84$_{.021}$  & 0.79$_{-.023}$ & 0.76$_{.002}$  \\
\quad +CO   & 0.81$_{.040}$    & 0.74$_{.093}$    & 0.63$_{-.008}$     & 0.71$_{.067}$    & 0.75$_{.012}$  & 0.83$_{.022}$  & 0.81$_{-.015}$ & 0.68$_{-.045}$ \\
\quad +FS    & 0.78$_{-.009}$   & 0.74$_{.041}$    & 0.63$_{.012}$      & 0.75$_{.004}$    & 0.77$_{-.010}$ &          -      &      -          &     -           \\
\quad +CO+FS & 0.82$_{.024}$    & 0.74$_{.040}$    & 0.64$_{.030}$      & 0.74$_{.054}$    & 0.77$_{.043}$  &          -      &         -       &         -       \\\midrule
\multicolumn{5}{l}{\textbf{Date}}                                                                         &                &                &                &                \\
\quad Base   & 0.75$_{.071}$    & 0.71$_{.073}$    & 0.58$_{-.005}$     & 0.69$_{.125}$    & 0.73$_{.076}$  & 0.82$_{-.001}$ & 0.81$_{-.004}$ & 0.75$_{.031}$  \\
\quad +CO   & 0.77$_{.095}$    & 0.71$_{.143}$    & 0.62$_{.024}$      & 0.69$_{.094}$    & 0.75$_{.039}$  & 0.81$_{.016}$  & 0.84$_{.017}$  & 0.66$_{.067}$  \\
\quad +FS    & 0.73$_{.093}$    & 0.71$_{.142}$    & 0.61$_{.131}$      & 0.65$_{.161}$    & 0.71$_{.029}$  &             -   &    -           &   -             \\
\quad +CO+FS & 0.76$_{.135}$    & 0.69$_{.113}$    & 0.62$_{.083}$      & 0.65$_{.180}$    & 0.73$_{.024}$  &           -     &         -       &         -       \\\midrule
\multicolumn{5}{l}{\textbf{Len.}}                                                                          &                &                &                &                \\
\quad Base   & 0.69$_{.062}$    & 0.62$_{.110}$    & 0.43$_{-.011}$     & 0.60$_{.039}$    & 0.66$_{.049}$  & 0.49$_{-.033}$ & 0.75$_{-.060}$ & 0.67$_{.050}$  \\
\quad +CO   & 0.76$_{.113}$    & 0.64$_{.130}$    & 0.55$_{.022}$      & 0.62$_{.109}$    & 0.67$_{.016}$  & 0.76$_{-.019}$ & 0.49$_{-.066}$ & 0.64$_{-.010}$ \\
\quad +FS    & 0.70$_{.059}$    & 0.61$_{.117}$    & 0.54$_{.028}$      & 0.62$_{.072}$    & 0.65$_{.074}$  &      -          &      -          &   -             \\
\quad +CO+FS & 0.76$_{.110}$    & 0.65$_{.078}$    & 0.58$_{.007}$      & 0.64$_{.063}$    & 0.69$_{.069}$  &          -      &         -       &   - \\\midrule            
\end{tabular}
}
\vspace*{3mm} 
\caption{Relative likability prediction results, with subscripts indicating improvements attributable to emotional ToM. ``CO'' denotes the inclusion of video context; ``FS'' signifies the usage of few-shot prompting with five samples; ``Rand.'' refers to random comment pairs; ``Vid.'' indicates comments from the same video; ``Date'' represents comments posted within a similar timeframe; ``Len.'' refers to similar comment lengths. }
\label{tab:popularity_like_prediction_main}
\end{table}

\begin{table*}[ht]
\small  
\centering
    \resizebox{\textwidth}{!}{%
\begin{tabular}{c|c|c|c|c|c|c}
\toprule
 & \multicolumn{1}{c|}{\textbf{Baseline} (Emotion)} & \multicolumn{1}{c|}{\textbf{Engaging} (Emotion)} & \multicolumn{1}{c|}{\textbf{ToM} (Intention)} & \multicolumn{1}{c|}{\textbf{Believe}}& \multicolumn{1}{c|}{\textbf{Denial}}& \multicolumn{1}{c}{\textbf{Extreme}} \\
\midrule
K & $\hat{\ell}_{\text{gen} | K}$ / $\hat{r}_{\text{gen} | K}$ / S. & $\hat{\ell}_{\text{gen} | K}$ / $\hat{r}_{\text{gen} | K}$ / S. & $\hat{\ell}_{\text{gen} | K}$ / $\hat{r}_{\text{gen} | K}$ / S. & $\hat{\ell}_{\text{gen} | K}$ / $\hat{r}_{\text{gen} | K}$ / S. & $\hat{\ell}_{\text{gen} | K}$ / $\hat{r}_{\text{gen} | K}$ / S. & $\hat{\ell}_{\text{gen} | K}$ / $\hat{r}_{\text{gen} | K}$ / S. \\
\midrule
1 & 2.20 / 0.25 / 0.89 & 7.25 / 0.59 / 0.79 & 2.26 / 0.39 / 0.72 & 3.23 / 0.42 / 0.77 & 1.91 / 0.38 / 0.85 & 2.37 / 0.41 / 0.77 \\
2 & 2.07 / 0.25 / 0.88 & 4.88 / 0.43 / 0.78 & 2.13 / 0.37 / 0.71 & 3.68 / 0.45 / 0.76 & 1.56 / 0.36 / 0.85 & 2.60 / 0.43 / 0.76 \\
3 & 2.51 / 0.28 / 0.88 & 4.90 / 0.47 / 0.78 & 2.20 / 0.39 / 0.71 & 3.42 / 0.46 / 0.75 & 1.84 / 0.38 / 0.84 & 2.88 / 0.45 / 0.75 \\
4 & 2.48 / 0.27 / 0.87 & 7.96 / 0.56 / 0.77 & 2.06 / 0.38 / 0.70 & 3.49 / 0.47 / 0.75 & 2.49 / 0.59 / 0.84 & 3.01 / 0.48 / 0.75 \\
5 & 2.58 / 0.29 / 0.87 & 7.44 / 0.56 / 0.77 & 2.00 / 0.37 / 0.70 & 3.30 / 0.47 / 0.74 & 2.48 / 0.56 / 0.84 & 2.83 / 0.49 / 0.74 \\
\bottomrule
\end{tabular}
}
\caption{Evaluation results for generated comments, where $\hat{\ell}_{\text{gen}|K}$ and $\hat{r}_{\text{gen}|K}$ denote estimated like and reply counts respectively based on K retrieved samples. S. represents the average similarity score among the K retrieved samples.}
\label{tab:gen_comment_eval}
\end{table*}

\section{Experimental Results}

\subsection{Engagement Modeling}

Figure~\ref{fig:persuasion_regression}(A) illustrates the effects of persuasion on video likes, suggesting that certain strategies consistently increase user engagement. 
The three most commonly used persuasion strategies—logical appeal (51\% of climate-related videos), authority (47\%) and common ground (36\%)—are each associated with a statistically lower audience engagement (see the distribution in Figure~\ref{fig:pipeline}). In contrast, emotional (33\%) and moral (26\%) appeals are linked to significantly higher levels of interaction, with morality-focused content emerging as the most effective strategy, producing an average increase of 2.1\% in video likes.

To examine temporal and format-specific trends, we disaggregate effects by video length. 
Figure~\ref{fig:persuasion_regression}(B) shows that the persuasive effect of moral rhetoric in short-form videos has increased over time, accompanied by a rise in authority-driven strategies. Here is an example of an authority appeal used in climate denialism, translated from Portuguese:
\begin{quote}
    \textit{``Climatologist Ricardo Felício stated in an interview that global warming is a hoax.''}
\end{quote}

These influences vary by publishing channels. The less effective trait, social norm, is perceived positively when videos are posted by international organizations.
Conversely, the overall effective trait, emotional appeals, does not extend to videos posted by scientific research institutes: 
\begin{quote}
    \textit{``Architect and urban planner Eduardo Pizarro, a CAJU member, showed how COVID-19 has most intensely affected the outskirts of the city of São Paulo.''}
\end{quote}

Although logical and statistical appeals generally receive lower levels of interaction, they tend to be more effective when delivered through national government channels. Below is an example with logical appeals:
\begin{quote}
    \textit{``If urgent measures are not taken, the planet's global temperature could rise by up to three degrees by the end of the 21st century. Therefore, ...''}
\end{quote}

Discussion topics further influence the effectiveness of informational appeals. Consistent with broader trends, sharing personal experiences and emphasizing moral values in DIY-related videos are associated with higher audience engagement, whereas logical and statistical narratives tend to reduce interaction. Although common ground generally has a negative impact, it can enhance engagement in sustainability-related themes:
\begin{quote}
    \textit{``It is important that we have a common understanding of what sustainability is—the balance between environmental, economic, and social issues.''}
\end{quote}

When videos address climate change in specific geo-locations, scarcity emerges as a strong predictor of comment engagement, even though it has minimal influence overall:
\begin{quote}
    \textit{``Sahara Desert was once a place with a huge forest, crisscrossed by rivers and lakes, and inhabited by a variety of animals.''} 
\end{quote}

Persuasion strategies also influence viewers' mental states.
Figure~\ref{fig:theory_of_mind}(A) shows correlations between 10 persuasion strategies and 7 ToM categories. 
Authority-based messaging tends to elicit stronger ToM responses (\textit{ps} $<$ .001 for all seven categories on two-tailed t-tests), whereas narratives centered on personal experiences generate fewer cognitively reflective comments.
Belief and emotion-related mental states exhibit greater variability compared to intention and percept, suggesting that they are more susceptible to change under different strategies.
Figure~\ref{fig:theory_of_mind}(B) further illustrates that emotion-oriented comments are more likely to be liked, while informative comments attract more replies.

\subsection{Popularity Prediction}


Table~\ref{tab:popularity_like_prediction_main} compares the accuracy of five LLMs, two encoder-based models, and a statistical approach across various experimental conditions.
GPT-4.1 achieves the best performance among LLMs (81\% accuracy) when provided with the video context. 
Surprisingly, BERTimbau, a Brazilian Portuguese encoder model, achieves 88\% accuracy even without contextual information, suggesting that the content of comments alone is often sufficient to predict engagement.
In addition, incorporating ToM mental states from user comments enhances the accuracy of engagement tendency predictions. 
Table~\ref{tab:popularity_like_prediction_main} shows the likability prediction gaps with and without emotional ToM narratives, showing that emotional ToM leads to an average improvement of 4.69\% in predictive performance.

\begin{figure}[t]
    \centering
    \includegraphics[width=0.6\linewidth]{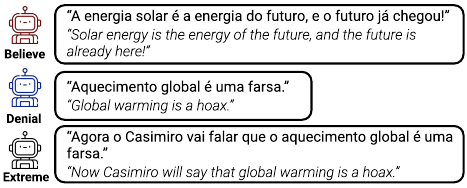}
    \caption{Sampled Portuguese comments generated by Believe, Denial, and Extreme models, with translations below.}
    \label{fig:comment_generation}
\end{figure}

\subsection{Comment Generation}

Fine-tuned LLMs show the ability to generate persuasive comments.
Table~\ref{tab:gen_comment_eval} presents the results of similarity-based proxy evaluations.
The \textit{Baseline (Emotion)} model, trained on randomly sampled comments, shows limited engagement effectiveness ($\hat{\ell}_{\text{gen} | 1} = 2.20$).
In contrast, the \textit{Engaging (Emotion)} model, fine-tuned on top-liked comments, achieves three times more engagement ($\hat{\ell}_{\text{gen} | 1} = 7.25$).
The model trained to generate comments with intentionality performs similarly to the baseline ($\hat{\ell}_{\text{gen} | 1} = 2.26$).

To assess the risks of opinion manipulation, we further examine belief-specific comment generation
The \textit{Denial} model ($\hat{\ell}_{\text{gen} | 1} = 1.91$) produces less engaging content than the \textit{Believe} model ($\hat{\ell}_{\text{gen} | 1} = 3.23$). However, when fine-tuned on extreme climate denial narratives, the \textit{Extreme} model ($\hat{\ell}_{\text{gen} | 1} = 2.37$) generates comments that potentially attract more likes and replies.
Figure~\ref{fig:comment_generation} shows representative generated samples from the three models, showing that the \textit{Extreme} model tends to include more details and rhetorical intensity, making its content more engaging.

\section{Discussions and Conclusions}

\subsection{Factors of User Engagement
}

We analyzed Brazilian YouTube videos on climate change and presented a detailed characterization of user engagement patterns using psycholinguistic traits. Our large-scale annotations reveal that content employing moral and emotional rhetoric consistently drives higher viewer engagement (Figure~\ref{fig:persuasion_regression}), a trend that also extended to comment interactions (Figure~\ref{fig:theory_of_mind}). These findings echo prior research on affective content~\cite{lerner2000beyond, han2020anger}.
Importantly, our data confirm the distinct role of content source: emotional expressions from research institute accounts received lower engagement compared to those from individual creators. This variation suggests the need to interpret engagement dynamics within the specific communicative contexts in which content is produced.

Our findings also diverge from conventional assumptions that facts and figures are the most effective tool for belief change~\cite{costello2024durably}. Although such strategies emphasizing statistical evidence on climate change may encourage initial engagement, especially when used by official government channels, its reach appears limited within YouTube’s short-form, algorithmically curated environment. As users increasingly favor short-form content, opportunities for disseminating detailed, fact-based information may have become constrained. These dynamics raise concerns about fact-checking efficiencies.

\subsection{Implications on Opinion Manipulation}

This study was motivated by concerns about the potential misuse of generative AI to manipulate public opinion. Specifically, we sought real-world examples of possibilities of such manipulation in climate conversations in the Global South. We presented case studies, one demonstrating that comment popularity can be accurately predicted and that psychological traits further enhance performance (Table~\ref{tab:popularity_like_prediction_main}); and another showing that persuasive comments can be generated using targeted profiles via fine-tuned LLMs (Table~\ref{tab:gen_comment_eval}). In a concerning example, a model fine-tuned on climate denialism content was capable of generating highly engaging comments that could potentially disseminate climate misinformation automatically and at scale (Figure~\ref{fig:comment_generation}).

The escalating volume of AI-generated content, coupled with its growing difficulty in human detection of such content~\cite{jakesch2023human,dugan2023real}, intensifies the risk of automated opinion manipulation~\cite{spitale2023ai}. Climate discourse is no exception to this risk, and this calls for the urgent need for governance around synthetic media. The rapid proliferation of short-form videos further complicates the challenge of verifying content authenticity (as shown in Figure~\ref{fig:short_long_distribution}), as swarms of synthetic narratives can be tailored to diverse user preferences and potentially create a misleading sense of consensus around socially sensitive issues~\cite{schroeder2025malicious}.

\subsection{Limitations}

This study has several limitations. It analyzed only textual content, leaving out multimodal elements that may influence persuasion. Engagement metrics did not account for factors such as recommendation algorithms or individual psychological differences. All findings are specific to Brazilian videos in Portuguese on YouTube.

\section*{Acknowledgments}

W. Dong and M.S. Locatelli contributed equally as co-first authors. 
V. Almeida and M. Cha are co-corresponding authors of this work.
This project was supported by the Microsoft Accelerate Foundation Models Research (AFMR) program.  
The authors thank Pedro Henrique Alves dos Santos, Raul Ferreira da Cruz Neto, and anonymous reviewers for their feedback.

\clearpage

\section*{Appendix -- A: Dataset Format}

The curated Brazilian YouTube climate discourse dataset is hosted on Zenodo.
We require all users to complete a formal access request to access this dataset.
The dataset is organized into two parquet files: 
\begin{itemize}
    \item \textbf{Video} (youtube\_video.parquet): YouTube video metadata together with ten persuasion, topic modeling, and channel classification labels. Each video is uniquely identified as \textsf{videoId}, totaling 226,775 videos.

    \item \textbf{Comment} (youtube\_comment.parquet): Top-level comments for all videos with one or more of the seven Theory-of-Mind labels. Each comment is uniquely identified as \textsf{id}, with 2,756,165 total comments. The attribute \textsf{videoId} can be matched to video metadata. Following the recommendations of previous work~\cite{liaw2023younicon}, we removed all user names and hashed user IDs to protect the identity of the users.
    
\end{itemize}

We include aggregated data that summarize comment attributes at the video level. This dataset is used for correlation analysis between 10 persuasion strategies and 7 categories of Theory of Mind (ToM) in case study 1.

\begin{itemize}
    \item \textbf{Aggregated YouTube} (aggregated\_youtube.parquet): These comments on videos include the average comment length, likes, replies, and 7 ToM categories for 88,983 videos with at least one comment.
    
\end{itemize}

\section*{Appendix -- B: Dataset Usage and Ethics}

We adhere to the dataset release protocols in~\cite{kirk2024prism} and outline the following instructions for YouTube content usage. This dataset is intended solely for research and educational purposes, including applications such as social media analysis and the evaluation of artificial intelligence models, including LLMs. The authors do not endorse any of the views or opinions expressed within the dataset.


YouTube video content and comments may include sensitive, biased, or potentially misleading information. We recommend that future use of this dataset apply appropriate filtering or content moderation to mitigate potential risks. In particular, we advise caution when using YouTube comment data to train generative models, as these models may inadvertently reflect or amplify biases present in the original content. Our experimental results indicate that fine-tuning language models on persuasive comments can lead to the reproduction of extreme viewpoints, such as climate change denial. We strongly recommend that any research utilizing this dataset includes an ethical considerations section that addresses potential societal and technical implications.

\section*{Appendix -- C: Related Datasets}
\label{appendix:datasets}

Prior to data collection, we conducted a non-exhaustive comparison of existing climate-related datasets. 
We summarize the comparison results, including the data source, intended usage and task, total dataset size and annotated sample size, and the language of the data.
\citet{choi2024climatemist} introduced the ClimateMist dataset for political stance and misinformation detection, comprising 146k tweets with 5k human annotations. Similarly, \citet{shiwakoti2024analyzing} expanded climate change text classification through the ClimaConvo dataset, with 15k tweets annotated for six key tasks: relevance detection, stance detection, hate speech identification, direction and target, and humor analysis. 
These datasets primarily focus on English-language content. Other studies have explored climate discourse in developing countries, such as Pirá~\cite{paschoal2021pira}, a dataset containing pt-br QA samples and MMT~\cite{dalal2023mmt}, which includes data in over 40 Indian languages. Our dataset occupies a unique intersection by bridging climate change discourse with psychological constructs such as persuasion and theory of mind (ToM), while also highlighting a country with significant influence in climate change discussions: Brazil.

\section*{Appendix -- D: Data Collection}

\begin{table}[h]
\small
\centering
\begin{tabular}{p{.94\linewidth}}
\toprule
\textbf{Brazilian Portugese Keywords}  \\
\midrule
mudança climática, aquecimento global, gases do efeito estufa, energia renovável, combustíveis fósseis, pegada de carbono, sustentável, ativismo climático, COP24, COP25, COP26, COP27, COP28, COP29, COP30, greenwashing, ecoansiedade, efeito estufa, acordo de Paris, ativista climático, aquecimento oceânico, nível do mar, emissão de gases, sequestro de carbono, desertificação, justiça climática, rarefação da camada de ozônio, mitigação climática, vulnerabilidade climática, adaptação climática, antropoceno, mudança ambiental, políticas climáticas, clima, reciclagem, poluição plástica, ação climática, meio ambiente, sem plástico, ecologia, zero desperdício, resíduos plásticos, acidificação oceânica, impostos de carbono, monitoramento ambiental, pacto ecológico europeu, recuperação verde, aquecimento global não existe, aquecimento global é falso, mudança climática não existe, mudança climática é mentira, resfriamento global, era glacial em 2030, farsa do aquecimento global, mentira climática, sem emergência climática, golpe do clima, manipulação do clima, ebulição global, parar o aquecimento global, salve o planeta, seja a mudança, emergência climática, salvar os oceanos, crise climática \\
\toprule
\textbf{English Translation} \\
\midrule
climate change, global warming, greenhouse gases, renewable energy, fossil fuels, carbon footprint, sustainable, climate activism, COP24, COP25, COP26, COP27, COP28, COP29, COP30, greenwashing, eco-anxiety, greenhouse effect, Paris Agreement, climate activist, ocean warming, sea level, gas emissions, carbon sequestration, desertification, climate justice, ozone layer depletion, climate mitigation, climate vulnerability, climate adaptation, anthropocene, environmental change, climate policies, climate, recycling, plastic pollution, climate action, environment, plastic free, ecology, zero waste, plastic waste, ocean acidification, carbon taxes, environmental monitoring, European Green Deal, green recovery, global warming doesn't exist, global warming is fake, climate change doesn't exist, climate change is fake, global cooling, glacial era 2030, global warming hoax, climate lie, no climate emergency, climate scam, climate manipulation, global boiling, stop global warming, save the planet, be the change, climate emergency, save the oceans, climate crisis \\
 
\bottomrule
\end{tabular}
\vspace*{3mm}
\caption{List of Brazilian Portuguese keywords used as queries to collect YouTube videos.}
\label{tab:keywords}
\end{table}

We curate a comprehensive list of 65 keywords relevant to the Brazilian context from the literature~\cite{baltasar2022analysis,salmi2022mudanccas} and reputable climate change glossaries~\cite{glossary-un,glossary-rsbp}.
Table~\ref{tab:keywords} shows the full list of keywords. 
We used YouTube Data API v3 to collect video metadata from the \texttt{search list} endpoint based on highest relevance for each week in the studied period. 
We set ``relevance language'' as Portuguese and ``region code'' as Brazil to maximize the context relevance when querying the curated list of keywords.
However, this does not guarantee that the results will meet these constraints. After the initial data collection process, we retrieved over 1.5 million videos of varying languages (including duplicates across different queries). We used fasttext~\cite{joulin2016fasttext} language identification to filter our dataset, resulting in 450k videos. We also removed videos and comments with invalid values (e.g. missing title or text) and duplicates.
To remove irrelevant content collected because of multiple meanings of keywords (e.g., ``climate'' in Portuguese can refer to a psychological atmosphere), we assessed their relevance to climate discussions using \emph{GPT-4.1-mini-2025-04-14} with a temperature value of 0 (validated using 300 manually annotated samples; F1 = 0.90, accuracy = 0.90).  This reduced the size of the dataset to the final size of 227k videos. We then collected the transcripts for each of these videos using the unofficial \texttt{youtube-transcript-api}. Finally, we separate the videos into short and long-form based on their duration, with short videos being those which are shorter than 3 minutes. We present descriptive statistics for our dataset in Table~\ref{tab:statistics}. 

\begin{table*}[tb]
    \centering
    \small
    \resizebox{\textwidth}{!}{%
    \begin{tabular}{lrrrrrrr}\toprule
                     & \textbf{Videos} & \textbf{Mdn. Duration} & \textbf{Mdn. Transcript Len.} & \textbf{Mdn. Likes} & \textbf{Total Comments} & \textbf{Channels} & \textbf{  Commenters} \\ \midrule
    \textbf{Short}   & 98,821         & 1.0                                & 148.0                                       & 5                                             & 695,730                 & 42,158                   & 462,909                    \\
    \textbf{Long}    & 127,954        & 12.7                               & 1,753.0                                     & 11                                            & 2,059,401               & 50,460                   & 1,194,181                  \\
    \textbf{All} & 226,775        & 4.0                                & 631.5                                     & 8                                             & 2,756,165               & 84,096                   & 1,527,855     \\ \bottomrule            
    \end{tabular} 
    }
    \caption{Statistics for our dataset. Duration is measured in minutes and transcript length in words. Mdn. refers to the median.}
    \label{tab:statistics}
\end{table*}

\begin{table}[t]
\centering
\small
\begin{tabular}{lcl}
\toprule
\multicolumn{3}{l}{\textbf{Topic}} \\
\midrule
Climate change & 20673 & \percentbar{9.1} \\
\rowcolor{gray!10} DIY crafts & 16950 & \percentbar{7.5} \\
Environment & 12704 & \percentbar{5.6} \\
\rowcolor{gray!10} Brazil & 11885 & \percentbar{5.2} \\
Sustainability & 9529 & \percentbar{4.2} \\
\rowcolor{gray!10} Renewable energy & 8390 & \percentbar{3.7} \\
Biology and ecology & 7999 & \percentbar{3.5} \\
\rowcolor{gray!10} Gas emissions and pollution & 7599 & \percentbar{3.4} \\
Solar energy & 5959 & \percentbar{2.6} \\
\rowcolor{gray!10} Agriculture and food & 5696 & \percentbar{2.5} \\
Fossil fuel & 4714 & \percentbar{2.1} \\
\rowcolor{gray!10} Amazon rainforest & 4369 & \percentbar{1.9} \\
Seas and oceans & 2962 & \percentbar{1.3} \\
\rowcolor{gray!10} Planet earth & 2690 & \percentbar{1.2} \\
Fresh water & 2365 & \percentbar{1.0} \\
\rowcolor{gray!10} COP summits & 2051 & \percentbar{0.9} \\
Paris agreement & 1990 & \percentbar{0.9} \\
\rowcolor{gray!10} Music & 1922 & \percentbar{0.8} \\
Animals and PET & 1591 & \percentbar{0.7} \\
\rowcolor{gray!10} Religion & 1360 & \percentbar{0.6} \\
Forests & 1344 & \percentbar{0.6} \\
\rowcolor{gray!10} Greenwashing and activism & 1339 & \percentbar{0.6} \\
Ice melting & 1311 & \percentbar{0.6} \\
\rowcolor{gray!10} Be the change & 1282 & \percentbar{0.6} \\
Africa & 1042 & \percentbar{0.46} \\
\bottomrule
\end{tabular}
\vspace*{3mm}
\caption{Topic modeling distributions.}
\label{tab:topic_modeling_distribution}
\end{table}

\section*{Appendix -- E: Topic Modeling}

\begin{figure}[ht]
    \centering
    \includegraphics[width=.8\linewidth]{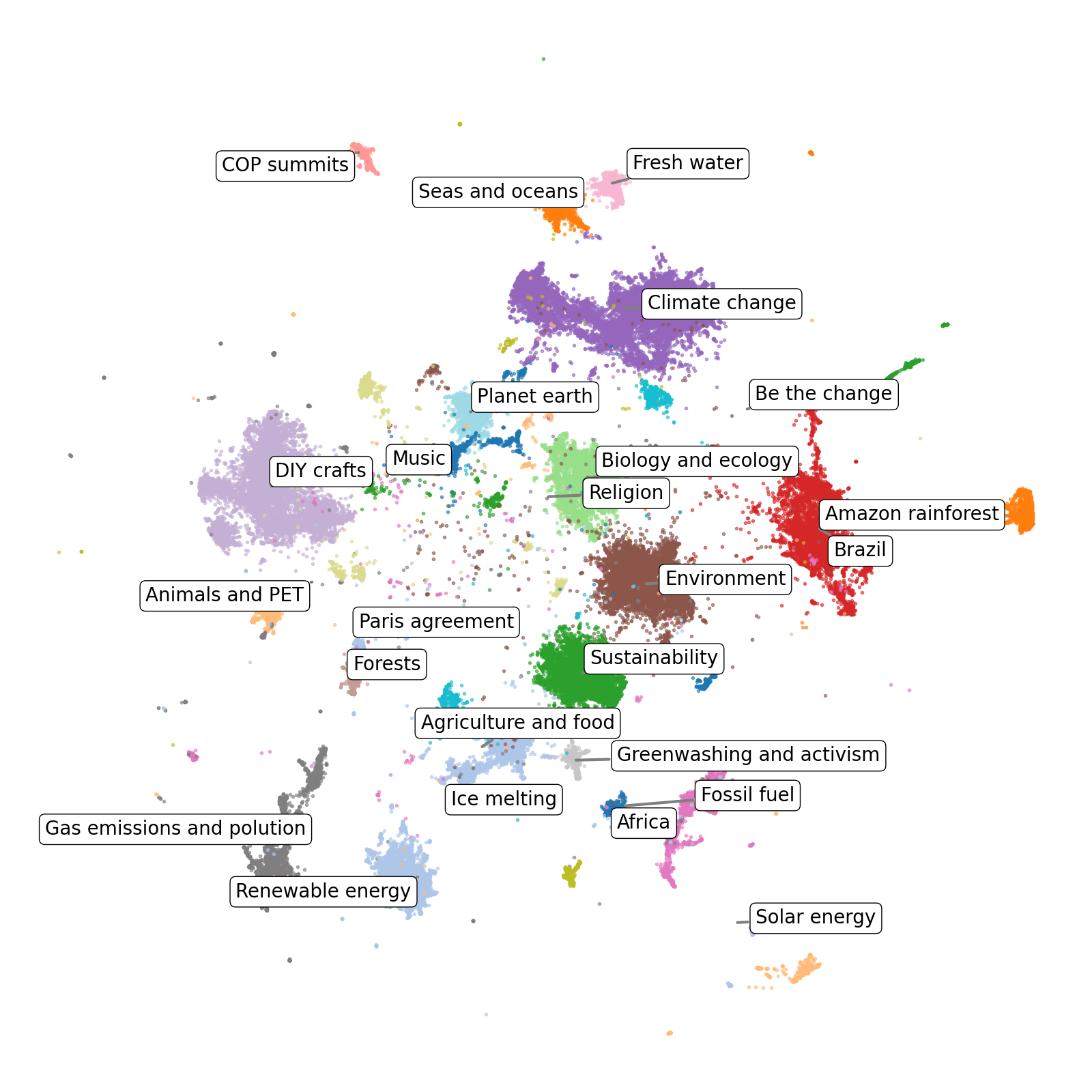}
    \caption{Topics generated for the videos in our dataset. The video embedding representations are reduced reduced to 2 using UMAP for visualization purposes.}
    \label{fig:topics}
\end{figure}

The topics in the dataset were obtained using BERTopic, which generates topic clusters and representative keywords from the video titles, which are then used to manually label each topic with a relevant, context-appropriate name. 
We use the pretrained model sentence-transformers/paraphrase-multilingual-MiniLM-L12-v2~\cite{reimers-2019-sentence-bert}, which has a maximum sequence length of 128, embedding size of 384 and supports over 50 languages, including Portuguese.
Then we use UMAP to reduce dimensionality with min dist=0; n components=5; n neighbors=15; and metric=`cosine'. Finally, we run HDBSCAN algorithm with min cluster size=1000 to capture overarching themes of climate discussions.
In total, we extracted 25 distinct topics in the dataset, ranging from social movements (``Be the change''), to forests and the Amazon, to sustainability, showcasing the diversity of the collected dataset. Figure~\ref{fig:topics} shows the distribution of the extracted topics in a 2 dimensional space, while Table~\ref{tab:topic_modeling_distribution} shows the full list of identified topics and their sample sizes. 

\section*{Appendix -- F: Channel Labeling}


YouTube is composed of many different types of users who produce content. In our dataset, we identified channels representing the government (e.g. prefectures, ministries), important organizations (e.g. the UN) and a plethora of other types of content creators. We used \emph{GPT-4.1-2025-04-14} with temperature set to 0 to label the channels into 15 pre-defined relevant categories decided upon manual verification of our dataset.
We validate these results using a set of 100 manually annotated samples (3 annotators, Fleiss' kappa = 0.64, LLM accuracy = 0.8). Table~\ref{tab:channel_distribution} shows the distribution of channel categories for the videos in our dataset.

\begin{table}[t]
\centering
\small
\begin{tabular}{lcl}
\toprule
\multicolumn{3}{l}{\textbf{Video Channel}} \\
\midrule
Individual content creators & 87080 & \percentbar{38.4} \\
\rowcolor{gray!10} Commercial companies & 24138 & \percentbar{10.6} \\
Traditional news outlets & 19903 & \percentbar{8.8} \\
\rowcolor{gray!10} Individual educators & 18597 & \percentbar{8.2} \\
Independent digital media & 16572 & \percentbar{7.3} \\
\rowcolor{gray!10} Non-profit organizations & 14001 & \percentbar{6.2} \\
Formal education organizations & 13360 & \percentbar{5.9} \\
\rowcolor{gray!10} Research institutes & 6253 & \percentbar{2.8} \\
Local government & 5684 & \percentbar{2.5} \\
\rowcolor{gray!10} National government & 4861 & \percentbar{2.1} \\
Industry representatives & 4801 & \percentbar{2.1} \\
\rowcolor{gray!10} Religious or spiritual organizations & 3899 & \percentbar{1.7} \\
International organizations & 1821 & \percentbar{0.8} \\
\rowcolor{gray!10} Political party & 1820 & \percentbar{0.8} \\
\bottomrule
\end{tabular}
\vspace*{3mm}
\caption{Distribution of videos per channel label.}
\label{tab:channel_distribution}
\end{table}

\section*{Appendix -- G: Persuasion Modeling}

\begin{figure*}[t]
    \centering
    \includegraphics[width=\linewidth]{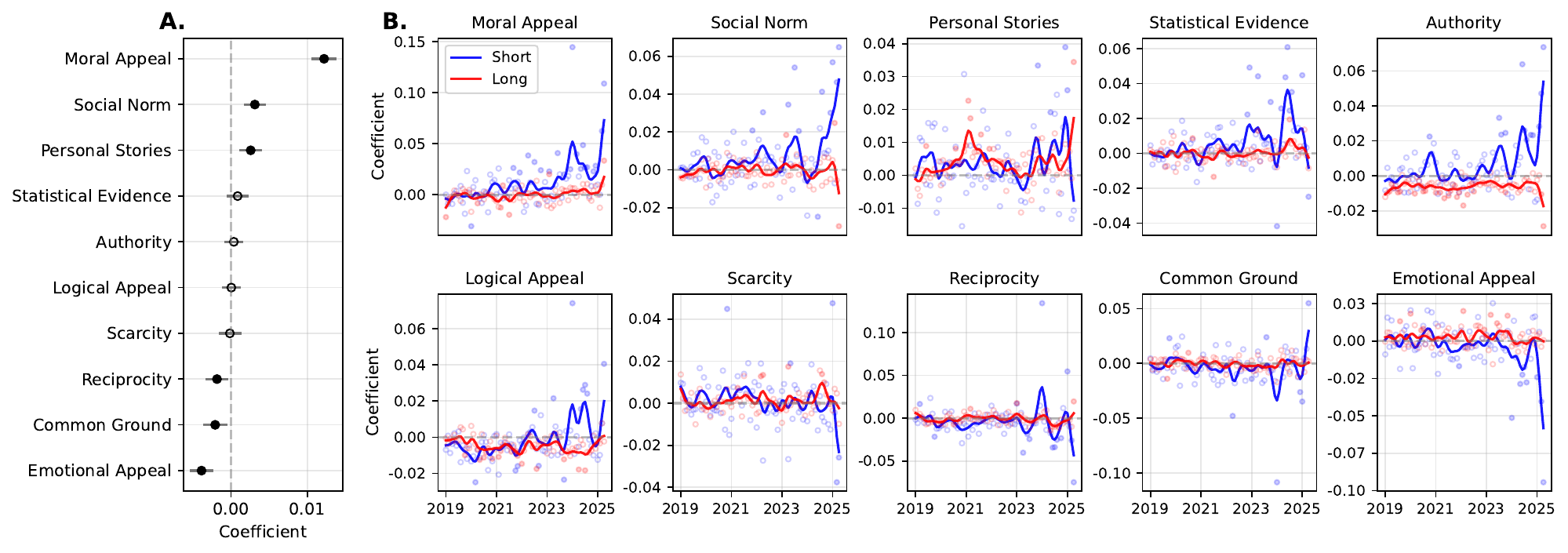}
    \caption{Effect of different persuasion strategies on the comment ratio for (A) all videos and (B) monthly trends by video length. Solid points indicate statistically significant regression coefficients at the 0.05 level.}
    \label{fig:persuasion_comment_ratio}
\end{figure*}

To assess the reliability of the persuasion results, two of the authors annotated 100 random samples and resolved any disagreements through discussion, until consensus was reached. 
The high level of agreement (cohen's $\kappa$=0.91) among two human and LLM annotations indicates the annotation reliability.
Accuracy scores were particularly high for personal stories (0.99), statistical evidence (0.98), emotional appeal (0.93), reciprocity (0.93), scarcity (0.92) and social norm (0.92). The rest of the categories still showed moderately high accuracy: common ground (0.87), moral appeal (0.86), authority (0.82), and logical appeal (0.81). Table~\ref{tab:validation} shows additional validation results for the persuasion annotation.

In addition to the persuasion strategy effects on video like ratios, we provide additional result on video comment ratios. Figure~\ref{fig:persuasion_comment_ratio} presents persuasion strategies effects on video comment ratios, where morality-emphasized content consistently receive higher engagements, and a notable increase on authority information is observed among short-form videos.

We further conducted negative binomial regressions to predict like and comment counts while controlling for views. Moral and emotional appeals were consistently associated with higher engagement, whereas authority and logical appeals show negative associations.
Table~\ref{tab:appendix_nb_regression} summarizes persuasion strategy influences on video like counts with Bonferroni-corrected p-values.

\begin{table}[h]
\centering
\begin{tabular}{lcccc}
\toprule
\textbf{Strategy} & \textbf{Coef.} & \textbf{95\% CI} & \textbf{z-score} \\
\midrule
Logical           & -0.0637  & [-0.07, -0.06] & -14.11\textsuperscript{***}    \\
Emotional         &  0.1456  & [0.13, 0.16]   & 24.96\textsuperscript{***}     \\
Statistical     & -0.0339  & [-0.04, -0.02] & -6.64\textsuperscript{***}     \\
Social Norm              & -0.1361  & [-0.15, -0.13] & -25.77\textsuperscript{***}    \\
Authority                & -0.1391  & [-0.15, -0.13] & -30.67\textsuperscript{***}    \\
Personal Stories         &  0.1013  & [0.09, 0.11]   & 18.64\textsuperscript{***}     \\
Moral Appeal             &  0.0869  & [0.08, 0.10]   & 14.04\textsuperscript{***}     \\
Reciprocity              & -0.0486  & [-0.06, -0.04] & -8.63\textsuperscript{***}     \\
Scarcity                 & -0.0052  & [-0.02, 0.01]  & -0.93          \\
Common Ground            & -0.0109  & [-0.02, 0.01]  & -1.85          \\
\bottomrule
\end{tabular}
\vspace*{3mm}
\caption{Influence of persuasion strategies on engagement metrics based on like counts using negative binomial regression.}
\label{tab:appendix_nb_regression}
\end{table}

\section*{Appendix -- H: Theory of Mind Modeling}

We validated the theory of mind (ToM) annotations using two methods. 
First, we randomly sampled 1000 comments and re-labeled them with GPT-4.1-2025-04-14, a larger model that is designed to support a context window of up to 1 million tokens~\cite{openai2025gpt41}.
Results suggest that the annotation results from the smaller model, GPT-4.1-mini, are aligned with GPT-4.1  while reducing latency and cost by 83\%, with the annotations being the same for 96\% of the comments for beliefs, 97\% for intentions, 95\% for desires, 92\% for emotions, 97\% for knowledge, 97\% for percepts, and 93\% for non-literal. This suggests that smaller models can process content at the level that is comparable to those of larger models in capturing nuanced mental states, making them suitable for large-scale comment annotations.

Second, we also conducted human labeling. Two of the authors independently labeled 100 randomly sampled comments and resolved any disagreements through discussion. 
The high level of agreement (cohen's $\kappa$=0.93) among two human and LLM annotations indicates the annotation reliability.
GPT-4.1-mini demonstrated strong alignment with human labels, achieving high accuracy across categories: 86\% for beliefs, 89\% for intentions, 91\% for desires, 81\% for emotions, 89\% for knowledge, 93\% for percepts, and 94\% for non-literal expressions. Table \ref{tab:validation} shows additional validation results for the ToM annotation.

\begin{table}[t]
    \centering
    \begin{tabular}{lrrr} \toprule
       \textbf{Task}  & $\mathbf{\kappa}$ &\textbf{F1} & \textbf{Aggreement \%/ Accuracy} \\ \midrule
       \textbf{Persuasion}  & 0.61	& 0.93 & 98\% \\
       \textbf{ToM}  &0.65 &	0.66	 & 83\%\\
       \bottomrule
         
    \end{tabular}
    \vspace*{3mm}
    \caption{Mean human validation results for persuasion and ToM using GPT-4.1. The Fleiss's $\kappa$ was calculated for the agreement between the 2 human annotators and the LLM.}
    \label{tab:validation}
\end{table}

\section*{Appendix -- I: Popularity Prediction}

\begin{table}[t]
\small
\centering
\begin{tabular}{lllllllll} \toprule
                          & \textbf{GPT-4.1} & \textbf{o4-mini} & \textbf{LLama-3.1} & \textbf{Phi-4}  & \textbf{LLama-4}         & \textbf{BERTimbau}       & \textbf{DeBerta V3}      & \textbf{Bradley–Terry}   \\ \midrule
\multicolumn{9}{l}{\textbf{Rand}}                                                                                                                                              \\
\quad Base & 0.75$_{\pm.02}$  & 0.69$_{\pm.03}$  & 0.57$_{\pm.07}$    & 0.72$_{\pm.03}$ & 0.68$_{\pm.02}$ & 0.86$_{\pm.05}$ & 0.82$_{\pm.02}$ & 0.69$_{\pm.04}$ \\
\quad + CO & 0.76$_{\pm.05}$  & 0.73$_{\pm.04}$  & 0.55$_{\pm.02}$    & 0.68$_{\pm.04}$ & 0.70$_{\pm.03}$ & 0.86$_{\pm.02}$ & 0.84$_{\pm.02}$ & 0.67$_{\pm.06}$ \\ \midrule
\multicolumn{9}{l}{\textbf{Vid.}}                                                                                                                                              \\
\quad Base & 0.77$_{\pm.05}$  & 0.73$_{\pm.03}$  & 0.59$_{\pm.04}$    & 0.74$_{\pm.07}$ & 0.69$_{\pm.03}$ & 0.82$_{\pm.05}$ & 0.77$_{\pm.04}$ & 0.68$_{\pm.04}$ \\
\quad + CO & 0.81$_{\pm.08}$  & 0.74$_{\pm.04}$  & 0.63$_{\pm.06}$    & 0.75$_{\pm.06}$ & 0.71$_{\pm.05}$ & 0.75$_{\pm.14}$ & 0.82$_{\pm.03}$ & 0.65$_{\pm.10}$ \\ \midrule
\multicolumn{9}{l}{\textbf{Date}}                                                                                                                                              \\
\quad Base & 0.75$_{\pm.03}$  & 0.71$_{\pm.04}$  & 0.58$_{\pm.02}$    & 0.73$_{\pm.03}$ & 0.69$_{\pm.02}$ & 0.82$_{\pm.02}$ & 0.80$_{\pm.05}$ & 0.69$_{\pm.05}$ \\
\quad + CO & 0.77$_{\pm.05}$  & 0.71$_{\pm.03}$  & 0.62$_{\pm.05}$    & 0.75$_{\pm.05}$ & 0.69$_{\pm.04}$ & 0.80$_{\pm.04}$ & 0.82$_{\pm.02}$ & 0.61$_{\pm.04}$ \\ \midrule
\multicolumn{9}{l}{\textbf{Len.}}                                                                                                                                              \\
\quad Base & 0.69$_{\pm.03}$  & 0.62$_{\pm.04}$  & 0.43$_{\pm.06}$    & 0.66$_{\pm.03}$ & 0.60$_{\pm.04}$ & 0.71$_{\pm.14}$ & 0.72$_{\pm.08}$ & 0.63$_{\pm.04}$ \\
\quad + CO & 0.76$_{\pm.03}$  & 0.64$_{\pm.03}$  & 0.55$_{\pm.06}$    & 0.67$_{\pm.04}$ & 0.62$_{\pm.03}$ & 0.59$_{\pm.17}$ & 0.76$_{\pm.03}$ & 0.60$_{\pm.05}$ \\ \midrule
\end{tabular}
\vspace*{3mm}
\caption{Relative likability prediction results, with subscripts indicating the 95\% confidence interval for the accuracy. ``CO'' denotes the inclusion of video context; ``Rand.'' refers to random comment pairs; ``Vid.'' indicates comments from the same video; ``Date'' represents comments posted within a similar timeframe; ``Len.'' refers to similar comment lengths.}
\label{tab:popularity_ci}
\end{table}

We set up different experimental conditions to include context when predicting comment popularity.
For each setting (\textbf{Rand.}, \textbf{Vid.}, \textbf{Date}, \textbf{Len.}), we randomly select 1k out of the 5k most popular comments in our dataset and match them to 1k low popularity comments depending on the conditions. Here popularity is defined by $\ell_k$, where k is a comment. $C^{pop}$ is the set of popular comments. The settings are thus defined as follows:
\begin{itemize}
    \item \textbf{Rand.} Sample 1000 random comments and pair them to the existing comments in $C^{pop}$, randomly.
    \item \textbf{Vid.} For each comment in $C^{pop}$, sample an unpopular comment from the same video.
    \item \textbf{Date} For each comment in $C^{pop}$, sample an unpopular comment from the same video that was published within 24 hours of the popular comment.
    \item \textbf{Len.} For each comment in $C^{pop}$, sample an unpopular comment from the same video that has at most 3 more (or less) words.      
\end{itemize}

We access the GPT-4.1, o4-mini, Phi-4, and Llama-4 models through the Microsoft Azure API. Specifically, we use the following versions for each model:
GPT-4.1 (2025-04-14)~\cite{openai2025gpt41},
Phi-4 (Version 7)~\cite{abdin2024phi},
o4-mini (2025-04-16)~\cite{o4mini}, and 
Llama-4-Maverick-17B-128E-Instruct (Version 1)~\cite{meta2025llama4}.
We use the Llama-3.1-8B-instruct~\cite{grattafiori2024llama} model available through huggingface locally.
For all these models, we employ the same prompt to assess the models' predictions. Additionally, we set the temperature=0 (or 0.001 for Llama models), frequency penalty=0, presence penalty=0 and top\_p=0.95 to reduce the variation of the experiments. 
For models that did not have all these parameters, we set only the ones that are applicable.
 For robustness, we also present replication results using a 5-fold validation strategy in Table~\ref{tab:popularity_ci}, which shows the average across all folds and 95\% CI for the zero-shot settings.

As for the training details for pre-trained language models, we fine-tune the BERTimbau and the DeBerta V3 models on an Nvidia A100 GPU 40 GB, with 128 GB memory and Intel Xeon Platinum 8360Y cpu, using an AdamW optimizer with $lr=5\times10^{-5}$ and the default Huggingface trainer parameters for 5 epochs using 5000 samples and a train-validation split of 0.8/0.2 and use the best model on the validation set for the test predictions.

\section*{Appendix -- G: Comment Generation}

For fine-tuning the models, we employ the LLaMa-factory framework~\cite{zheng2024llamafactory} and use the default parameters to fine-tune each of the comment generation models on a Nvidia A100 GPU 40 GB, with 128 GB memory and Intel Xeon Platinum 8360Y cpu, using LoRa with rank r=8. For each model, we used 4000 training samples and 1000 validation samples, and trained for 3 epochs with a learning rate of $lr=10^{-4}$ and precision of BF16.

For generating comments, we use a set of 100 video titles and descriptions not included in the training set as inputs and applied the following temperature values, [0, 0.1, 0.2, ..., 1], with 100 experiments for each temperature condition, totaling 1100 inference results for each model.
As for the evaluations, we obtain embeddings of generated comments using paraphrase-multilingual-MiniLM-L12-v2. We use Facebook AI Similarity Search (FAISS)~\cite{douze2024faiss} to retrieve $K$ most similar comments from the reference dataset as the proxy of likability and reply tendency of generated comments.

\bibliographystyle{unsrtnat}
\bibliography{0_main}  

\end{document}